\documentstyle[12pt]{article}
\def\gmmu{\gamma _{\mu}}
\def\gmnu{\gamma_{\nu}}

\def\gmf{\gamma _{5}}

\def\ll{\langle }
\def\rl{ \rangle }
\def\nableft{ \stackrel{\leftarrow}{\nabla}}
\def\nabright{ \stackrel{\rightarrow}{\nabla}}
\newcommand{\beq}{\begin{equation}}
\newcommand{\eeq}{\end{equation}}
\newcommand{\bea}{\begin{eqnarray}}
\newcommand{\eea}{\end{eqnarray}}

\begin{document}
\renewcommand{\thefootnote}{\fnsymbol{footnote}}
                                        \begin{titlepage}
\begin{flushright}
TECHNION-PHYS-94-16 \\
hep-ph/9411422
\end{flushright}
\vskip1.8cm
\begin{center}
{\LARGE
Soft gluon suppression of $ 1/N_{c} $ contributions in color
suppressed heavy meson decays
            \\ }
\vskip1.5cm
 {\Large Igor~Halperin}
 \\
\vskip0.2cm
       Technion - Israel Institute of Technology   \\
       Department of Physics  \\
       Haifa, 32000,  Israel \\
{\small e-mail address: higor@techunix.technion.ac.il}\\

\vskip1.5cm
{\Large Abstract:\\}
\parbox[t]{\textwidth}{
We discuss the non-factorizable terms in color suppressed (Class II)
decays. Our emphasis is on the non-perturbative soft gluon exchange
mechanism, which has been previously found to be responsible for the
rule of dicarding $ 1/N_{c} $ in the Class I decays. The non-factorizable
contribution to the decay $ \bar{B}^{0} \rightarrow D^{0} \pi^{0} $
at the tree level is estimated within the light cone QCD sum rule method
which combines the
technique of the QCD sum rules with the description of the pion in terms
of the set of wave functions of increasing twist. We find that the same
soft gluon exchange mechanism tends to cancel the $ 1/N_{c} $ term in
the factorized amplitude.
}

\vspace{1.0cm}
{\em submitted to Physics Letters B }
\end{center}
                                                \end{titlepage}

{\bf 1}. There has been raising interest during a last few years in testing
the
factorization approach to nonleptonic heavy meson decays. The current
activity in this direction  has
been triggered by the observation \cite{BSW}
that the available data for D-meson decays seems to support a rule of
discarding $ 1/N_{c} $ terms in factorized hadronic matrix elements
\cite{BGR}, which is in contrast to the standard prescription that keeps
such terms. This approximate cancellation of $ 1/N_{c} $ terms in the
D decays has been explicitly checked \cite{BS87} within the QCD sum rule
approach.
Surprisely, the recent data on B-meson decays \cite{CLEO}
signals that discarding $ 1/N_{c} $ terms is not a universal rule for
all channels. The $ 1/N_{c} $-suppression has rather a dynamical character
and varies for different channels [5-10]. Still, there exist (both
perturbative and non-perturbative) indications
that patterns of deviation from the factorization approximation are
alike inside each separate class of decays if they are
classified according to
factorization properties of corresponding effective Hamiltonians.
In particular, it has been found in Ref. \cite{BS},\cite{BSF} that the
$1/N_{c} $ rule is likely to hold in the Class I decays (see (5) below).
The effect has a dynamical origin, and is due to non-perturbative gluon
effects which have been estimated in \cite{BS},\cite{BSF},\cite{KR} by the
QCD sum rule
method. The present letter is aimed to test an importance of these
effects on the so-called color suppressed (Class II) decays. As a
particular example, we will study the decay $ \bar{B}^{0} \rightarrow
D^{0} \pi^{0} $ following the method suggested in Ref. \cite{BS}.

At the tree level, weak hadronic Cabibbo favored B-decays correspond to
the quark transitions $ b \rightarrow c \bar{c}s $ and $ b \rightarrow
c \bar{u} d $ and are governed by the effective Hamiltonian
\beq
H_{eff} = \frac{G_{F}}{\sqrt{2}}[ V_{cb}V_{cs}^{*} ( C_{1}(\mu) O_{1}^{c} +
C_{2}(\mu) O_{2}^{c}) + V_{cb}V_{ud}^{*} ( C_{1}(\mu) O_{1}^{u} + C_{2} (\mu)
O_{2}^{u}) ] \eeq
where ( $ \Gamma_{\mu} = \gmmu ( 1 - \gmf) $ )
\beq
O_{1}^{u} = (\bar{c} \Gamma_{\mu} b ) ( \bar{d} \Gamma_{\mu} u ) \; \; and
\; \; O_{2}^{u} = ( \bar{d}\Gamma_{\mu} b )( \bar{c} \Gamma_{\mu} u)
\eeq
and the operators $ O_{1,2}^{c} $ are obtained from $ O_{1,2}^{u} $ by
the substitution $ (\bar{d}, u) \rightarrow ( \bar{s}, c) $. The Wilson
coefficients $ C_{i}( \mu) $ are due to the renormalization of  the bare
Hamiltonian $ H_{W} \sim O_{1} $ by hard gluons with virtualities larger
than $ \mu^{2} = O( m_{b}^{2} ) $. In the leading-log approximation
$ C_{1,2} = \frac{1}{2} ( C_{+} \pm C_{-} ) $ with \cite{AM}
\beq
C_{\pm} (\mu) = \left( \frac{ \alpha_{s}(\mu)}{ \alpha_{s}(M_{W})}
\right)^{\frac{6 \gamma_{\pm}}{ 33 - 2 n_{fl}}}
\eeq
with $ \gamma_{-} = - 2 \gamma_{+} = 2   $ . At $ \mu \simeq 5 \; GeV $
and $
n_{fl} = 5 $, this yields ( $ \Lambda_{\bar{MS}} \simeq 200 \; MeV $ )
\beq
C_{1} = 1.117 \; \; and \; \; C_{2} = - 0.266
\eeq
Within the factorization approximation one can distinguish between three
classes of decays for which the corresponding amplitudes have the
following structure \cite{BSW}
\beq
A_{I} \sim a_{1}(\mu) <O_{1}> , A_{II} \sim a_{2}(\mu) <O_{2}> , A_{III}
\sim [ a_{1}(\mu) + x a_{2}(\mu) ] <O_{1}>
\eeq
Here $ <O_{i}> $ are the (factorized) hadronic matrix elements of
the operators $ O_{i} $ and $ a_{i}(\mu) $ are QCD factors related to the
coefficients $ C_{i}(\mu) $ :
\beq
a_{1}(\mu) = C_{1}(\mu ) + \frac{1}{N_{c}} C_{2}(\mu) \; \; , \; \;
a_{2}(\mu) = C_{2}(\mu) + \frac{1}{N_{c}} C_{1}(\mu)
\eeq
An attempt of the global fit of nonleptonic B-decays yields \cite{GKKP},
\cite{BHP} \beq
a_{1} = 1.05 \pm 0.10 \; \; and \; \; a_{2} = 0.25 \pm 0.05
\eeq
It should be mentioned that the very possibility of the global fit is
rather questionable in view of both experimental and theoretical
uncertainties and the expected variation between the naive factorization
and the $ 1/N_{c} $ rule for different channels [5-10]. Still,
note that while the value of $ a_{1} $ is consistent with suggestive
discarding the $ 1/N_{c} $ correction in (6), this is not the case for
$ a_{2} $. The problem of the (non-)factorization thus becomes particularly
important for the decays falling into the second class
in the above scheme. These are called color suppressed decays in view of
the fact that they are non-zero only due to the QCD-induced operator $
O_{2}$. To illustrate the basic difference between the class I and class II
amplitudes, consider the decays $ \bar{B}^{0} \rightarrow D^{+} \pi^{-} $
and $ \bar{B}^{0} \rightarrow D^{0} \pi^{0} $. For the former decay, the
operator $ O_{1} $ gives the leading contribution as $ N_{c} \rightarrow
\infty $ :
\beq
\langle D^{+} \pi^{-} | O_{1} | \bar{B}^{0} \rangle
= \langle \pi| \bar{d} \Gamma_{\mu} u | 0 \rangle \langle D^{+} | \bar{c}
\Gamma_{\mu} b | \bar{B} \rangle + O ( \frac{1}{N_{c}^{2}} ) \; ,
\eeq
while $ O_{2} $ is factorized only after applying the Fiertz
transformation
\beq
O_{2} = \frac{1}{N_{c}} O_{1} + 2 ( \bar{c}\Gamma_{\mu} t^{a} b ) ( \bar{d}
\Gamma_{\mu}t^{a} u ) \equiv \frac{1}{N_{c}} O_{1} + \tilde{O}_{2}
\eeq
Note that the matrix element $ \langle D^{+}\pi^{-} | \tilde{O}_{2} | \bar{B}
\rangle $ vanishes in the factorization approximation because of the
color conservation. An appealing method to estimate this nonfactorizable
contribution has been proposed in \cite{BS}. There it has been shown that
the above matrix element can be reduced to a simpler one $ ( p_{\alpha}^{(B)}
- p_{\alpha}^{(D)} ) \langle D | \bar{c} \Gamma_{\mu} g \tilde{G}_{\alpha
\mu} b | B \rangle $ by virtue of the short distance operator product
expansion (OPE) , while the latter matrix element is fixed by the heavy
flavor symmetry and can be expressed via the experimental number $
M_{B_{*}}^{2} - M_{B}^{2} \simeq 0.46 \; GeV^{2} $. Consequently, the
non-factorizable piece is of the same order as the factorizable $ (1/N_{c} )
O_{1} $ part of $ O_{2} $, and has the opposite sign. This observation
justifies the (approximate) rule of discarding $ 1/N_{c} $ corrections in
the class I decays.

Let us now consider the $ \bar{B}^{0} \rightarrow D^{0} \pi^{0} $ decay.
Then $ O_{2} $ factorizes and
\bea
\langle D^{0} \pi^{0} | H_{eff} | \bar{B}^{0} \rangle \sim ( C_{2} + \frac{
C_{1}}{N_{c}} ) \langle D| \bar{c} \Gamma_{\mu} u | 0 \rangle \langle \pi
| \bar{d} \Gamma_{\mu}b | B \rangle \nonumber \\
 + C_{1} \langle D \pi | 2 ( \bar{c} \Gamma_{\mu} t^{a} u ) ( \bar{d}
\Gamma_{\mu} t^{a} b ) | B \rangle
\eea
Our task is to estimate the non-factorizable contribution to this decay
which is given by the second matrix element in (10). Note that
within the
approach of Ref. \cite{BS}, this contribution will be proportional to the
matrix element $ p_{\alpha} \langle \pi  | \bar{d} \Gamma_{\mu}
\tilde{G}_{\alpha \mu} b | B \rangle $ (see below)
, which is not known from general
principles and deserves a separate calculation. We will present below an
estimate for $ \langle D^{0} \pi^{0} | \tilde{O}_{1} | \bar{B}^{0}
\rangle $  based on the so-called light cone QCD sum rules method (
see \cite{BH} , \cite{BBKR} and references therein) which
combines the
traditional technique of the QCD sum rules \cite{SVZ} with a description
of an emitted light particle ( $ \pi^{0} $ in our case ) in terms of the
wave functions of increasing twists. We will argue that the suppression
of $ 1/N_{c} $ contributions via the soft gluon exchange mechanism \cite{BS}
holds also for $ \bar{B}^{0} \rightarrow D^{0} \pi^{0} $ and probably for
other Class II decays. However, for the color suppressed
decays, such cancellation is not what is welcomed by the phenomenology if
one assumes the validity of the global fit (7).
We will come back to this point later on.

{\bf 2}. To implement  the two-step strategy
suggested in Ref. \cite{BS}, we start with the correlation function
\beq
A_{\mu} = i \int dx \, e^{ipx} \ll \pi^{0} | \bar{u}(x)\gmmu \gmf c(x) ,
\tilde{O}_{1}(0) |\bar{B}^{0} \rangle = i f_{D} p_{\mu} \langle
D \pi | \tilde{O}_{1}
| \bar{B} \rangle \frac{1}{m_{D}^{2} - p^{2}}  + ...
\eeq
where the ellipses stand for higher states contributions. At large Euclidean
$ p^{2} $ the correlation function (11) can be calculated in QCD. The leading
contribution is due to the soft gluon emission from the quark loop.
A simple calculation yields the following "sum rule" \cite{BSF}
\beq
if_{D} \ll D \pi | \tilde{O}_{1} | \bar{B} \rl  \frac{1}{m_{D}^{2} -
p^{2}} = \frac{1}{4 \pi^{2}} p_{\mu}
\ll \pi | \bar{d}g \tilde{G}_{\mu \nu} \gmnu \gmf b | \bar{B} \rl \left[
\frac{1}{ - p^{2}} - \frac{m_{c}^{2}}{(-p^{2})^{2}} \ln{\frac{m_{c}^{2} -
p^{2}}{m_{c}^{2}}} \right] + ...
\eeq
which is to be satisfied in the duality interval $ 1 \; GeV^2 < - p^{2} <
4 \; GeV^2 $. (In obtaining (12), we have omitted a contribution
proportional to
the matrix element $  p_{\mu} \ll \pi | \bar{d} gG_{\mu
\nu} \gmnu b | \bar{B} \rl \simeq
p_{0} \ll \pi | \bar{d} gG_{0 i} \gamma_{i}
b| \bar{B} \rl $ as it is down by the inverse heavy quark mass : $ \gamma_{i}
b = O(\frac{1}{m_{b}})$ ). As the ratio of the kinematic factors remains
approximately constant and equal 1 in the duality region, we obtain
\beq
\ll D^{0}(p) \pi^{0}(q)| \tilde{O}_{1}| \bar{B}^{0}(p+q) \rl \simeq
- \frac{i}{4 \pi^{2} f_{D}} p_{\mu} \ll \pi(q)| \bar{d} g \tilde{G}_{\mu \nu}
\gmnu \gmf b | \bar{B}(p+q) \rl
\eeq
(Note that in the case at hand the use of the short distance OPE is
justified by the fact that the final c-quark is heavy and its velocity is
small in the rest frame of the b-quark \cite{BS} ).
It is important to point out that the formula (13) is obtained at the
zero order in $ \alpha_{s} $, and cannot be considered as the
renormalization group covariant one. The answer (13) refers to the
normalization point $ \mu = O(m_{b}) $, the entire $ \mu $-dependence
being implicit. At the one-loop level the operator $ \tilde{O}_{1} $ mixes
with $ O_{2} $, while the operator $ \bar{d} g \tilde{G}_{\mu \nu} \gmnu
\gmf b $ mixes with the operator $ m_{b}^{2} \bar{d} \gmmu b $ and
operators vanishing on the equations of motion :
\newpage
\bea
( \bar{d} g \tilde{G}_{\mu \nu} \gmnu \gmf b)^{ \mu_{2}^{2} } &=& (1 -
\frac{4}{3} ( N_{c} - \frac{1}{N_{c}} ) \frac{\alpha_{s}}{4 \pi} \ln{ \frac{
\mu_{2}^{2}}{\mu_{1}^{2}}} ) ( \bar{d} g \tilde{G}_{\mu \nu} \gmnu \gmf
b)^{\mu_{1}^{2}} \nonumber \\
&+& \frac{C_{F}}{2} \frac{\alpha_{s}}{4 \pi}
\ln{\frac{\mu_{2}^{2}}{\mu_{1}^{2}}} ( - \frac{2}{3}  \bar{d} (
\hat{\nableft^{2}} \gmmu + \gmmu \hat{\nabright^{2}} ) b  \\
&+& \frac{2}{3}  \bar{d} (  \hat{\nableft} \nabla_{\mu} +
\nabla_{\mu} \hat{\nabright} ) b  - 2  m_{b} \bar{d} \sigma_{\mu \nu}
\nabright_{\nu} b - 2 i  m_{b}^{2} \bar{d} \gmmu b )^{\mu_{1}^{2}}
\nonumber
\eea
where we have set $ m_{d} = 0 $ and $ C_{F} = \frac{N_{c}^{2} - 1}{ 2 N_{c}}
= 4/3 $. Thus, one has to understand (13) as the tree level relation.
For consistency, in what follows we will omit one-loop contributions
  altogether.

To find the new matrix
element (13), consider another correlation function
\bea
T_{\alpha}(p,q) &=& i \int
dx \, e^{ipx} \ll \pi(q) |
\bar{d}(x)g\tilde{G}_{\alpha \mu}\gmmu \gmf b(x) \bar{b}(0) i \gmf d(0)|0
\rl  \nonumber \\
                &=& \frac{m_{B}^{2}f_{B}}{m_{b}} \frac{1}{m_{B}^{2} -
(p+q)^{2}} \, [ p_{\alpha} f_{1}(p^{2}) + ( p_{\alpha} + 2 q_{\alpha})
f_{2}(p^{2}) ]
\eea
where $ \ll B | \bar{b} i \gmf d | 0 \rl = m_{B}^{2} f_{B} /m_{b} $ .
At this stage the light cone QCD sum rules method suggests a simple
and straightforward way of the calculation. At
large Euclidean $  (p+q)^{2} $ the leading contribution to (15) is
\beq
T_{\alpha}(p,q) = i g \int \frac{d^{4}x d^{4}k}{(2 \pi)^{4} ( m_{b}^{2}
- k^{2})} e^{i(p-k)x} \langle \pi | \bar{d}(x) \tilde{G}_{\alpha \mu}(x)
\gmmu \gmf ( \hat{k} + m_{b}) \gmf d(0) | 0 \rangle
\eeq
(here $ \hat{k} \equiv k_{\mu} \gamma_{\mu} $), that can be further
evaluated introducing the pion wave function (WF)
of twist 3 $ \phi_{3 \pi} $
\bea
\langle \pi | \bar{d}(x) g
G_{\mu \nu} (vx) \sigma_{\alpha \beta} \gmf d(0)
| 0 \rangle = - \frac{i}{\sqrt{2}} f_{3 \pi} [ q_{\alpha} ( q_{\mu}
g_{\nu \beta} - q_{\nu} g_{\mu \beta}) - q_{\beta} ( q_{\mu} g_{\nu
\alpha} - q_{\nu} g_{\mu \alpha})]  \nonumber \\
\times \int D \,\alpha_{i} \phi_{3 \pi}( \alpha_{i}) e^{iqx( \alpha_{1}
+ v \alpha_{3}) }
\eea
( here $ D \alpha_{i} = d \alpha_{1} d \alpha_{2} d \alpha_{3}
\delta( \alpha_{1} + \alpha_{2} + \alpha_{3} - 1) $ and $ f_{3 \pi}
\simeq 0.0035 \; GeV^{2} $ for $  \mu^{2} \simeq 1 \; GeV^{2} $ \cite{CZ} )
, and the set of WF's of twist 4 : \bea
\langle \pi^{0} | \bar{d}(x) \gmmu i g \tilde{G}_{\alpha \beta} (vx)
d(0) | 0 \rangle = - \frac{f_{\pi}}{\sqrt{2}} \left[
q_{\beta} (  g_{\alpha \mu}
- \frac{x_{\alpha} q_{\mu}}{ qx} ) - q_{\alpha} ( g_{\beta \mu} -
\frac{x_{\beta} q_{\mu} }{ qx} ) \right] \nonumber \\
\times \int D \alpha_{i} \tilde{\phi}_{\perp}
(\alpha_{i} e^{ iqx ( \alpha_{1} + v \alpha_{3} ) } \nonumber \\
- \frac{f_{\pi}}{\sqrt{2}} \frac{q_{\mu}}{ qx} ( q_{\alpha} x_{\beta}
- q_{\beta} x_{\alpha} ) \int D \alpha_{i} \tilde{\phi}_{\parallel}
(\alpha_{i}) e^{iqx( \alpha_{1} + v \alpha_{3}) }
\eea
Two more WF's of twist 4 $ \phi_{\perp} \; , \;
\phi_{\parallel} $ are defined analogously to (17) with the
substitution $ ( i g \tilde{G}_{\alpha \beta} ) \rightarrow ( \gmf g
G_{ \alpha \beta} ) $ ).

Then, using the identity
\beq
\int_{0}^{1} du \int D \alpha_{i}\delta( u - \alpha_{1} - \alpha_{3} )
\Phi ( \alpha_{i} ) = \int_{0}^{1} du \int_{0}^{u} d \alpha_{3}
\Phi( \alpha_{1} = u - \alpha_{3} , \alpha_{2} = \bar{u} , \alpha_{3}) \; ,
\eeq
the answer for the simplest tree diagram can be written as
\beq
T_{\alpha} = \frac{q_{\alpha}}{\sqrt{2}} \int_{0}^{1} \frac{du }{ m_{b}^{2}
- ( p +
uq)^{2}} \int_{0}^{u} d \alpha_{3} [ - 2 (pq) f_{3 \pi}  \phi_{3 \pi}
 + f_{\pi} m_{b} ( \tilde{\phi}_{\parallel} - 2 \tilde{\phi}_{\perp}) ]
(u - \alpha_{3}, \bar{u} , \alpha_{3})
\eeq
A systematic study of higher twist WF's beyond the asymptotic regime has
been done in Ref. \cite{BF2}. This analysis has been based on the
expansion in representations of the so-called collinear conformal group
SO(2,1), which is a subgroup of the full conformal group SO(4,2) acting
on the light cone. The asymptotic WF's are defined as contributions of
operators with the lowest conformal spin and unambigouosly fixed by the
group structure. Pre-asymptotic corrections correspond to the operators
with the next-to-leading conformal spin, whose numerical values have been
calculated by the QCD sum rules method. The result for $ \phi_{3 \pi} $
reads \cite{BF2} :
\bea
\phi_{3 \pi} ( \alpha_{i}) &=& 360 \alpha_{1} \alpha_{2} \alpha_{3}^{2}
[ 1 + \omega_{1,0} \frac{1}{2} ( 7 \alpha_{3} - 3) + \omega_{2,0} (
2 - 4 \alpha_{1} \alpha_{2} - 8 \alpha_{3} + 8 \alpha_{3}^{2}) \nonumber \\
                           &+& \omega_{1,1}
( 3 \alpha_{1} \alpha_{2} - 2 \alpha_{3} + 3 \alpha_{3}^{2} ) + ... ]
\eea
where $ \omega_{1,0} = -2.88 \; , \; \omega_{2,0} = 10.5 \; , \;
\omega_{1,1} = 0 $ in a low normalization point \cite{CZ},\cite{BF2}. For
the twist 4 WF's the relevant first order formulas are
\bea
\tilde{\phi}_{\perp} (\alpha_{i})   &=& 30 \delta^{2} ( 1 -
\alpha_{3})
\alpha_{3}^{2} [ \frac{1}{3} + 2 \varepsilon ( 1 - 2 \alpha_{3}) ]
\nonumber \\
\tilde{\phi}_{\parallel} (\alpha_{i}) &=& - 120 \delta^{2}
 \alpha_{1} \alpha_{2} \alpha_{3} [ \frac{1}{3} + \varepsilon ( 1 - 3
\alpha_{3}) ]
\eea
where the parameter $ \delta^{2} \simeq 0.2 \; GeV^{2} $ ( at $ \mu^{2}
\simeq 1 \; GeV^{2} $ ) is defined via
\beq
\ll 0 | \bar{d} g \tilde{G}_{\mu \nu} \gmnu \gmf u| \pi^{+} \rl =
i \delta^{2} f_{\pi} q_{\mu}
\eeq
and $ \varepsilon \simeq 0.5 $ is the weight of the first conformal spin
correction. We will need the above set of parameters renormalized to a
higher normalization point $ \mu^{2} \simeq \mu_{b}^{2} = \sqrt{m_{B}^{2}
-m_{b}^{2}} \simeq (2.4 \; GeV)^{2} $ which is the characteristic
virtuality of the b-quark in the B-meson \cite{CZ90},\cite{BBKR}. The
corresponding anomalous dimensions can be found in \cite{BF2}. The results
read \cite{BBKR}  $ f_{3 \pi} = 0.0026 \; GeV^{2} , \;
\omega_{1,0} = -2.18 , \; \omega_{1,1} = - 2.59 , \; \omega_{2,0} = 8.12
, \; \delta^{2} \simeq 0.18 , \; \varepsilon \simeq 0.4 $.

To match the answer (20) with the B-meson contribution to the correlation
function (15), we note that (20) can be re-written as the dispersion
integral with the expression $ (m_{b}^{2} - \bar{u} p^{2})/u $ being the
mass of the intermediate state. The duality prescription tells that
this invariant mass has to be restricted from above by the duality threshold
$ s_{0} \simeq 35 \; GeV^{2} $ (this value is obtained from corresponding
two-point sum rules). As it is easy to see, this transforms into the
effective cut-off from below in the u-integral \cite{BKR},\cite{BH}. Finally,
we make the standard Borel transformation suppressing both higher states
resonances and higher Fock states in the full pion wave function.
Under the Borel transformation $ -(p+q)^{2} \rightarrow M^{2} $
\bea
\frac{1}{m_{B}^{2} - (p+q)^{2}} \rightarrow \exp{(-
\frac{m_{B}^{2}}{M^{2}}) } \nonumber \\
\frac{1}{ m_{b}^{2} - ( p+uq)^{2}} \rightarrow \frac{1}{u} \exp{( -
\frac{m_{b}^{2} - \bar{u}p^{2}}{ u M^{2}})}
\eea
Our final sum rule takes the form
\bea
f_{1}(p^{2}) &=& - f_{2}(p^{2}) = - \frac{1}{\sqrt{2}} \frac{m_{b}}{2
m_{B}^{2} f_{B}}
\int_{0}^{1} du \int_{0}^{u} d
\alpha_{3} \; \exp{( \frac{m_{B}^{2}}{M^{2}} - \frac{m_{b}^{2} - \bar{u}
p^{2}}{u
M^{2}}) } \Theta(u - \frac{m_{b}^{2} - p^{2}}{ s_{0} - p^{2}}) \nonumber \\
& \times & \left[ - f_{3 \pi} \frac{m_{b}^{2} - p^{2}}{u^{2}} \phi_{ 3 \pi}
+ f_{\pi} \frac{m_{b}}{u} ( \tilde{\phi}_{\parallel} - 2
\tilde{\phi}_{\perp}) \right] ( \alpha_{i})
\eea

{\bf 3}. Now we turn to numerical estimates. In evaluating (25), we have
used
the following set of parameters : $ m_{b} = 4.7 \; GeV , \; m_{B} = 5.28
\; GeV , \; s_{0} \simeq 35 \; GeV^{2} , \; f_{B} \simeq 135 \; MeV $
\cite{BKR},\cite{BBKR}. The Borel mass $ M^{2} $ has been varied in the
interval from 8 to 20 $ GeV^{2} $. We have found that within the
variation of $M^{2}$ in this region, the result changes no more that by
10 \% and yields for $ p^{2} \simeq m_{D}^{2} $
\beq
f_{1}(m_{D}^{2}) \simeq - f_{2}(m_{D}^{2}) \simeq \frac{1}{\sqrt{2}} \times
0.08 \; GeV^{2}
\eeq
Then for the matrix element of interest we obtain
\beq
p_{\alpha} \ll \pi | \bar{d} \tilde{G}_{\alpha \mu} \gmmu \gmf b | B \rl
\simeq M_{D}^{2} f_{1}(m_{D}^{2}) + M_{B}^{2} f_{2} ( m_{D}^{2}) \simeq
- \frac{1}{\sqrt{2}} \times 1.9 \; GeV^{4}
\eeq
For the factorizable amplitude due to the operator $ (1/N_{c}) O_{2} $ we
have
\bea
M_{f} &=& \frac{1}{N_{c}} i f_{D} p_{\mu} \ll \pi(q)| \bar{d} \gmmu b|
B(p+q) \rl \nonumber \\
&=& \frac{i}{N_{c}} f_{D} p_{\mu} [ 2 q_{\mu} f_{\pi}^{+} (p^{2}) +
p_{\mu} ( f_{\pi}^{+} (p^{2} ) + f_{\pi}^{-}(p^{2}) ) ]
\eea
The value of the form factor $ f_{\pi}^{+} (m_{D}^{2}) $ can be read off
the results of Ref. \cite{BKR} where this quantity has been calculated
(for the charged pion) by virtue of the light cone QCD sum rules method.
The answer is
\beq
f_{\pi}^{+} (m_{D}^{2}) \simeq - \frac{1}{\sqrt{2}} \times 0.3
\eeq
where the number $ ( - 1/ \sqrt{2} ) $ is due to the different isospin
structure in our case. For the second form factor we use the model \cite{BSW}
\beq
f_{\pi}^{-} (p^{2}) = - f_{\pi}^{+}(p^{2}) \frac{m_{B} - m_{\pi}}{m_{B} +
m_{\pi}} \simeq - f_{\pi}^{+}(p^{2})
\eeq
and therefore neglect the second term in (28). In the nomenclature of
Ref. \cite{BS}, we thus obtain the following estimate for the ratio of
 the non-factorizable to the factorizable $ 1/N_{c} $
amplitudes :
\beq
r \equiv \frac{M_{nf}}{M_{f}} \simeq - \frac{N_{c}}{4 \pi^{2} f_{D}^{2}}
\frac{ p_{\alpha} \ll \pi | \bar{d} \tilde{G}_{\alpha \mu} \gmmu \gmf b |
B \rl }{ ( m_{B}^{2} - m_{D}^{2}) f_{\pi}^{+}(m_{D}^{2})} \simeq - 0.7
\eeq
where we have used the value $ f_{D} \simeq 170 \; MeV $ corresponding to
omitted $ \alpha_{s} $-corrections in the relevant two-point sum rule
\cite{BKR,BBKR}. Thus, our final result (31) suggests that the
non-factorizable contribution tends to cancel the factorizable $ 1/N_{c} $
amplitude due to the operator $ (1/N_{c}) O_{2} $ (see (10)), in
agreement with expectations of Ref. \cite{BS},\cite{BSF}. The sign of the
effect can also be compared with
the estimate  within the QCD sum rules method done in Ref.
\cite{KR} for the weak decay $ B \rightarrow J/ \Psi \; K $, which also
belongs to the class of color suppressed decays. The authors of \cite{KR}
have found that the power corrections due to the gluon condensate partly
cancel the $ 1/N_{c} $ factorizable amplitude, i.e. their sign of $ r $
is negative, too. On the other hand, within the approach of Ref.
\cite{BS} the
non-factorizable amplitude for this decay can be approximately
expressed via the matrix element $ p_{\alpha} \ll K | \bar{s}
\tilde{G}_{\alpha \mu} \gmmu \gmf b | B \rl $, which can be extracted from
(27) assuming the SU(3) limit. One has to bear in mind, however, that the
approach of Ref. \cite{BS} cannot be literally applied to the decay $ B
\rightarrow J/ \Psi \; K $. The poor stability of the corresponding
sum rule \cite{BSF} implies that perturbative corrections or operators of
higher dimensions must be important there.

{\bf 4}. In this letter we have estimated the soft gluon
exchange mechanism contribution to the non-factorizable amplitude of
 the $ \bar{B}^{0} \rightarrow D^{0} \pi^{0} $ decay and found the
tendency for the cancellation of $ 1/N_{c} $, in seeming contradiction
with (7). One should emphasize, however,
that our result does not mean the actual contradiction of the theory with the
available data \cite{CLEO}. One possible sourse of the disagreement may be
the unjustified use of the global fit leading to the particular
numbers (7). Such fit implies that the factorization properties are alike
in all non-leptonic two-body B-decays. The validity of this assumption
has to be examined in QCD.
Probably the more important origin of disagreement with (7) are large
 perturbative $O( \alpha_{s}) $-
corrections which are not taken into account in our calculation, and
correspond to different contributions to non-factorizable amplitudes.
Estimates made in Ref. \cite{KR} indicate that radiative corrections are
important for a complete evaluation of  non-factorizable amplitudes for
the Class II decays. As we have mentioned, in a calculation including
radiative corrections one needs an accurate separation of $
O(\alpha_{s}) $ corrections related to the matrix element itself and $
O(\alpha_{s}) $ terms due to the mixing with the two-particle operators, cf.
(14).

In the recent paper \cite{Buras} it has been stressed that
non-factorizable amplitudes are very important in one more aspect. The
point is that the coefficient $ a_{2} $ becomes strongly $ \mu $- and scheme-
 dependent beyond the leading-log approximation. Depending on the
renormalization scheme, $ a_{2} $ can scale crudely from 0.1 to 0.2.
This is in strong contrast with $ a_{1} $ which exhibits very weak
$ \mu -$ and scheme- dependence. The importance of higher order QCD
corrections for an accurate calculation of
 $ a_{2} $ can be intuitively understood as a consequence of the
fact that $ a_{2} $ is the difference of the nearly equal numbers,and thus
is very sensetive to their precise values. Since
the factorized amplitudes are $ \mu -$ and scheme-independent, only the
non-factorizable
contributions can remove this scheme (and $ \mu $- ) dependence in the
physical amplitudes. Unfortunately, we have nothing to add in respect
to this important problem in view of our neglecting perturbative $
O(\alpha_{s}) $  corrections. At the same time, the observation of Ref.
\cite{Buras} suggests that the perturbative corrections to the matrix
elements are presumably more important for the Class II decays than for
those of the Class I. The main question is whether the account for hard
gluon loops  is able to change the sign of the ratio (31).
A calculation of the matrix
element $ \ll D^{0} \pi^{0} | \tilde{O}_{1} | \bar{B}^{0} \rl $ including
the radiative corrections can be done either by the methods of Ref.
\cite{BS}, \cite{BS87,KR}, or directly by
the light cone QCD sum rules method \cite{BH}. In the latter approach, one
has to calculate
the three-point correlation function of the D- and B-meson currents
and the effective Hamiltonian between the vacuum and the pion states. We
have explicitly checked that in this case the leading $ O(\alpha_{s}^{0}
) $ contribution is again given by a combination of the three-particle
WF's of twists 3 and 4. However, the corresponding estimate based on
retaining only these terms manifests a poor stability, that indicates the
importance of radiative corrections or/and higher twist effects in the
corresponding sum rule. This instability does not occur in the above sum
rule, that formally justifies our neglecting the radiative and higher twist
correction.
Undoubtfully, the complete evaluation of
perturbative gluon effects on hadronic matrix elements of interest is
needed before any comparison  with the data. To our knowledge, this work is
currently in progress \cite{KR}.

I wish to thank Boris Blok for drawing my attention to this problem and
valuable discussions.


\clearpage
\begin{thebibliography}{99}

\bibitem{BSW}
M.~Bauer, B.~Stech and M.~Wirbel, Z.Phys. {\bf C34} (1987) 103.

\bibitem{BGR}
A.S.~Buras, J.M.~Gerald and R.~Ruckl,
Nucl. Phys. {\bf B268} (1986) 16.

\bibitem{BS87}
B.~Blok and M.A.~Shifman,
Sov. J. Nucl. Phys. {\bf 45} (1987) 135, 301, 522.

\bibitem{CLEO}
M.S.~Alam et.  al. (CLEO Collaboration), Phys. Rev. {\bf D50} (1994) 43.

\bibitem{BS}
B.~Blok and M.~Shifman, Nucl. Phys. {\bf B389} (1993) 534; {\bf B399} (1993)
441, 459.

\bibitem{BSF}
B.~Blok and M.A.~Shifman, in " Proceeding of the Fermilab meeting DPF 92",
ed. C.H.~Albright, P.H.~Kasper, R.~Raja and J.~Yon, World Scientific (1993).

\bibitem{KR}
A.~Khojamirian and R.~Ruckl , MPI -PhT/94-26 , LMU 05/94.

\bibitem{GKP}
M.~Gourdin, Y.Y.~Keum and X.Y.~Pham, PAP/LPTHE/94-32.

\bibitem{JS}
J.M.~Soares, TRI-PR-94-78.

\bibitem{Buras}
A.S.~Buras, MPI-PhT/94-60.

\bibitem{AM}
G.~Altarelli and L.Maiani, Phys. Lett. {\bf B52} (1974) 352; \\
M.K.~Gaillard and B.W.~Lee, Phys. Rev. Lett. {\bf 33} (1974) 108.

\bibitem{GKKP}
M.~Gourdin, A.N.~Kamal, Y.Y.~Keum and X.Y.Pham, Phys. Lett. {\bf B333}
(1994) 507.

\bibitem{BHP}
K.~Browder, K.~Honscheid and S.~Playfer, in "B Decays", ed. S.~Stone, World
Scientific (1994).

\bibitem{BH}
V.~Braun and I.~Halperin, Phys. Lett. {\bf B328} (1994) 457.

\bibitem{BBKR}
V.~Belyaev, V.~Braun, A.~Khodjamirian and R.~Ruckl, MPI-PhT/94-62,
CEBAF-TH-94-22, LMU 15/94.

\bibitem{SVZ}
M.I.~Shifman, A.I.~Vainshtein and V.I.~Zhakharov, Nucl. Phys. {\bf B147}
(1979) 385.

\bibitem{CZ}
V.L.~Chernyak and A.R.~Zhitnitsky, Phys. Rep. {\bf 112} (1984) 173.

\bibitem{CZ90}
V.L.~Chernyak and A.R.~Zhitnitsky, Nucl. Phys. {\bf B345} (1990) 137.

\bibitem{BF2}
V.M.~Braun and I.E.~Filyanov, Z. Phys. {\bf C48} (1990) 239.

\bibitem{BKR}
V.M.~Belyaev, A.Khodjamirian and R.Ruckl, Z. Phys. {\bf C60} (1993) 349.
\end{thebibliography}
\end{document}